\documentclass[12pt]{article}
\usepackage{graphicx}
\usepackage[retain-unity-mantissa=false, exponent-product=\cdot]{siunitx}
\DeclareSIUnit{\UCN}{UCN}
\usepackage{hyperref}
\usepackage{todonotes}
\usepackage{tikz}
\usepackage{tikzscale}
\usepackage{pgfplots}

\newcommand\pubnumber{CIPANP2018-Schreyer}
\newcommand\pubdate{\today}

\def\TRIUMF{TRIUMF\\4004 Wesbrook Mall, Vancouver, BC V6T 2A3, Canada}
\def\support{\footnote{Work supported by CFI, the Canada Research Chairs program, JSPS, KEK, NSERC, RCNP Osaka, Research Manitoba, TRIUMF, and University of Winnipeg.}}

\def\Title#1{\begin{center} {\Large #1 } \end{center}}
\def\Author#1{\begin{center}{ \sc #1} \end{center}}
\def\Address#1{\begin{center}{ \it #1} \end{center}}

\newcommand\pubblock{\rightline{\begin{tabular}{l} \pubnumber\\
         \pubdate  \end{tabular}}}
\newenvironment{Abstract}{\begin{quotation}  }{\end{quotation}}
\newenvironment{Presented}{\begin{quotation} \begin{center} 
             PRESENTED AT\end{center}\bigskip 
      \begin{center}\begin{large}}{\end{large}\end{center} \end{quotation}}

\begin{document}
\begin{titlepage}
\pubblock

\vfill
\Title{Towards TUCAN's Search for the Neutron Electric Dipole Moment}
\vfill
\Author{Wolfgang Schreyer\\on behalf of the TUCAN collaboration\support}
\Address{\TRIUMF}
\vfill
\begin{Abstract}
The TRIUMF ultracold advanced neutron (TUCAN) collaboration is a transpacific collaboration with the objective to measure the neutron electric dipole moment. We aim for an unprecedented sensitivity of \SI{1e-27}{\elementarycharge\centi\meter}, an improvement by a factor of 30 over the current upper limit for this elusive quantity. To achieve this goal, we are planning a next-generation source for ultracold neutrons, based on a new, dedicated neutron spallation source at TRIUMF, liquid deuterium as cold moderator, and superfluid helium as ultracold-neutron converter. To operate at the full beam power of \SI{20}{\kilo\watt}, a $^3$He fridge will provide a cooling power of \SI{10}{\watt} at \SI{0.8}{\kelvin}, cooling the converter to a temperature of \SI{1.1}{\kelvin}.

Thanks to extensive optimization of neutron moderation, heat transport in superfluid helium, and ultracold-neutron transport we expect a density of polarized ultracold neutrons in the experiment of \SI{350}{\per\cubic\centi\meter}, allowing us to reach the aspired statistical sensitivity in less than \num{400} beam days.
\end{Abstract}
\vfill
\begin{Presented}
Thirteenth Conference on the Intersections of Particle and Nuclear Physics\\
Palm Springs, CA, USA, May 29--June 3, 2018
\end{Presented}
\vfill
\end{titlepage}
\def\thefootnote{\fnsymbol{footnote}}
\setcounter{footnote}{0}

\section{Introduction}

How the matter-antimatter asymmetry in the universe came about is still one of the unsolved mysteries in physics and cosmology. As worked out by Sakharov~\cite{0038-5670-34-5-A08}, processes violating CP symmetry are a necessary ingredient for such an asymmetry to arise in the early universe.

Although CP violation can be observed within the Standard Model of particle physics in Kaon decays~\cite{PhysRevLett.13.138}, B-meson decays~\cite{PhysRevLett.87.091801}, and potentially in neutrino oscillations~\cite{PhysRevD.95.096014}, the effects are not large enough to explain the large matter-antimatter asymmetry observed today. To explain it, new sources of CP violation beyond the Standard Model are needed.

Electric dipole moments (EDMs) of leptons, atoms, and nucleons constitute a direct violation of CP symmetry and are some of the most sensitive probes for CP violation~\cite{POSPELOV2005119}. Measurements of the neutron EDM combining Ramsey's method of separated oscillating fields~\cite{PhysRev.78.695} with ultracold neutrons (UCNs) are one prominent example.

Ultracold neutrons have energies of a few hundred nanoelectronvolts and can be trapped for hundreds of seconds by bottles of certain materials, magnetic fields, and gravity. Additionally, they can be polarized to a high degree, making them uniquely suitable for Ramsey-type measurements. These measurements have reached impressive sensitivities, putting an upper limit on the neutron EDM of \SI{3.0e-26}{\elementarycharge\centi\meter}~\cite{PhysRevD.92.092003}. This limit is still far above the minuscule neutron EDM of \SI{1e-32}{\elementarycharge\centi\meter} predicted by the Standard Model, but it puts stringent constraints on many theories introducing new sources of CP violation and predicting a much larger neutron EDM~\cite{POSPELOV2005119}.

Progress of neutron-EDM experiments is currently limited by the small number of ultracold neutrons that the handful of sources worldwide can provide~\cite{PhysRevC.95.045503}, since accumulating the necessary statistical sensitivity takes years. A major avenue to reach higher sensitivity is to develop new, intense sources for ultracold neutrons.

Recent developments of new sources have focused on superthermal processes: a cold neutron scattering in a converter material can induce solid-state excitations and lose almost all of its energy, becoming an ultracold neutron~\cite{GOLUB1975133}. Cooling the converter to sufficiently low temperatures suppresses the inverse process of up-scattering and allows the ultracold neutrons to exit the converter and be guided along UCN-reflecting guides into an experiment.

Two converter materials are commonly used: solid deuterium~\cite{PhysRevC.97.012501,LAUSS201498,Kahlenberg2017} and superfluid helium~\cite{PhysRevC.90.015501,PhysRevLett.108.134801,Ahmed:2018fvb}. Solid deuterium cooled to a temperature of \SI{5}{\kelvin} offers a rich spectrum of solid-state excitations and high UCN-production rate, but the UCN density that can be extracted is limited by its high neutron-absorption cross section~\cite{Yu1986}. Conversely, superfluid helium has a lower UCN-production cross section, but by cooling it to temperatures around \SI{1}{\kelvin} and reducing the abundance of neutron-absorbing $^3$He isotopes, up-scattering and absorption can be highly suppressed~\cite{GOLUB1979387}.

To perform a measurement of the neutron EDM with a sensitivity of \SI{1e-27}{\elementarycharge\centi\meter}, the TUCAN collaboration is currently planning a next-generation source of ultracold neutrons using a neutron spallation source and a superfluid-helium converter. This source will provide a second experimental port that can be used by other ultracold-neutron experiments.

\section{Production and losses of ultracold neutrons in superfluid helium}

The dispersion relations of free neutrons and phonons in superfluid helium cross at an energy of \SI{1}{\milli\electronvolt}. A cold neutron with that energy can excite a single phonon, lose virtually all its energy, and get converted to an ultracold neutron. Detailed measurements of neutron scattering in superfluid helium show that multi-phonon scattering can contribute as well at energies up to \SI{5}{\milli\electronvolt}~\cite{PhysRevC.92.024004,KOROBKINA2002462}.

The UCN-loss rate in superfluid helium is given by
\begin{equation}
\tau^{-1} = \tau_\mathrm{wall}^{-1} + \tau_\mathrm{up}^{-1} + \tau_\mathrm{abs}^{-1} + \tau_\beta^{-1}.
\end{equation}

The wall-loss lifetime $\tau_\mathrm{wall}$ is typically tens to hundreds of seconds. The up-scattering rate is strongly temperature-dependent and roughly follows
\begin{equation}
\tau_\mathrm{up}^{-1} \approx B \cdot \left( \frac{T}{\SI{1}{\kelvin}} \right)^7
\end{equation}
with $B$ between \SIlist{0.008;0.016}{\per\second}~\cite{PhysRevC.93.025501}. The absorption lifetime $\tau_\mathrm{abs}$ is dominated by absorption on $^3$He but can be increased to more than \SI{1000}{\second} with isotopically purified helium. The storage lifetime is ultimately limited by the $\beta$-decay lifetime of free neutrons of $\tau_\beta = \SI{880.2+-1.0}{\second}$~\cite{PDG2018}.

To reduce the up-scattering rate to a similar level as the wall-loss rate, the superfluid helium has to be cooled to a temperature around \SI{1}{\kelvin}. With a UCN-production rate $P$, the total number of UCN accumulated in the source after the target has been irradiated for a time $t$ is
\begin{equation}
N = P \tau \left[ 1 - \exp\left(-\frac{t}{\tau}\right) \right],
\end{equation}
reaching $N(t \rightarrow \infty) = P \tau$ for irradiation times much longer than the storage lifetime.

\section{Neutron spallation and moderation}

\subsection{Beamline and spallation target}

In preparation for a new UCN source, we set up a dedicated beamline at TRIUMF \cite{BL1U}. TRIUMF's cyclotron provides a \SI{483}{\mega\electronvolt} proton beam of which up to \SI{40}{\micro\ampere} can be diverted onto the new tungsten spallation target. With a recently installed prototype UCN source~\cite{Ahmed:2018fvb}---originally developed in Japan~\cite{PhysRevLett.108.134801}---the beamline is currently limited to \SI{1}{\micro\ampere}. To increase UCN production by a factor of \num{40}, the planned next-generation source will be designed to operate at the full beam power of \SI{20}{\kilo\watt}.

\subsection{Moderator optimization}

\begin{figure}
    \centering
    \includegraphics[width=0.5\textwidth]{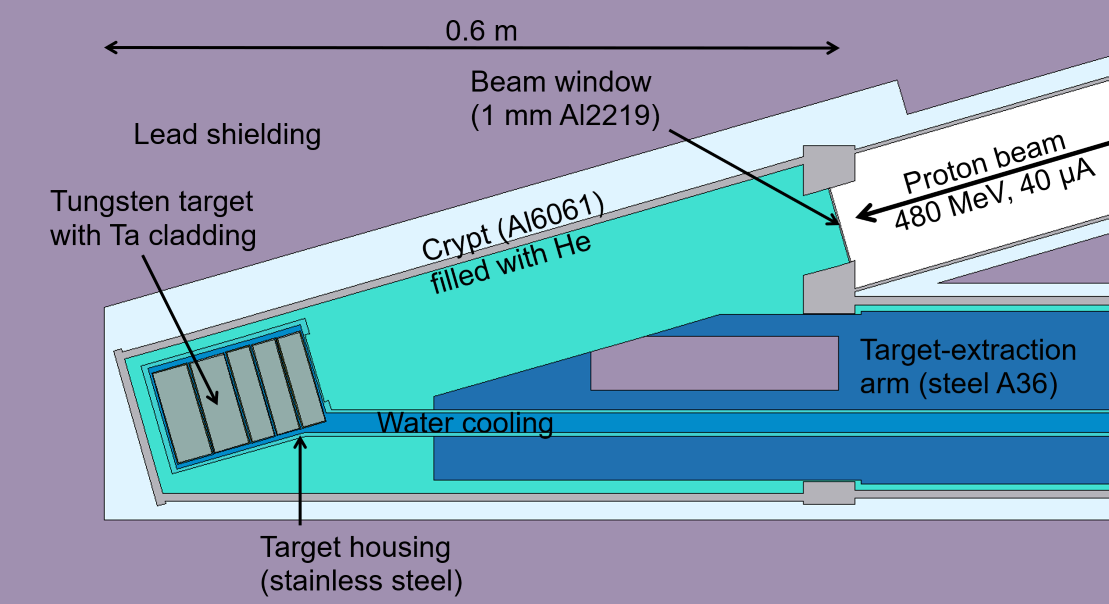}
    \includegraphics[width=0.45\textwidth]{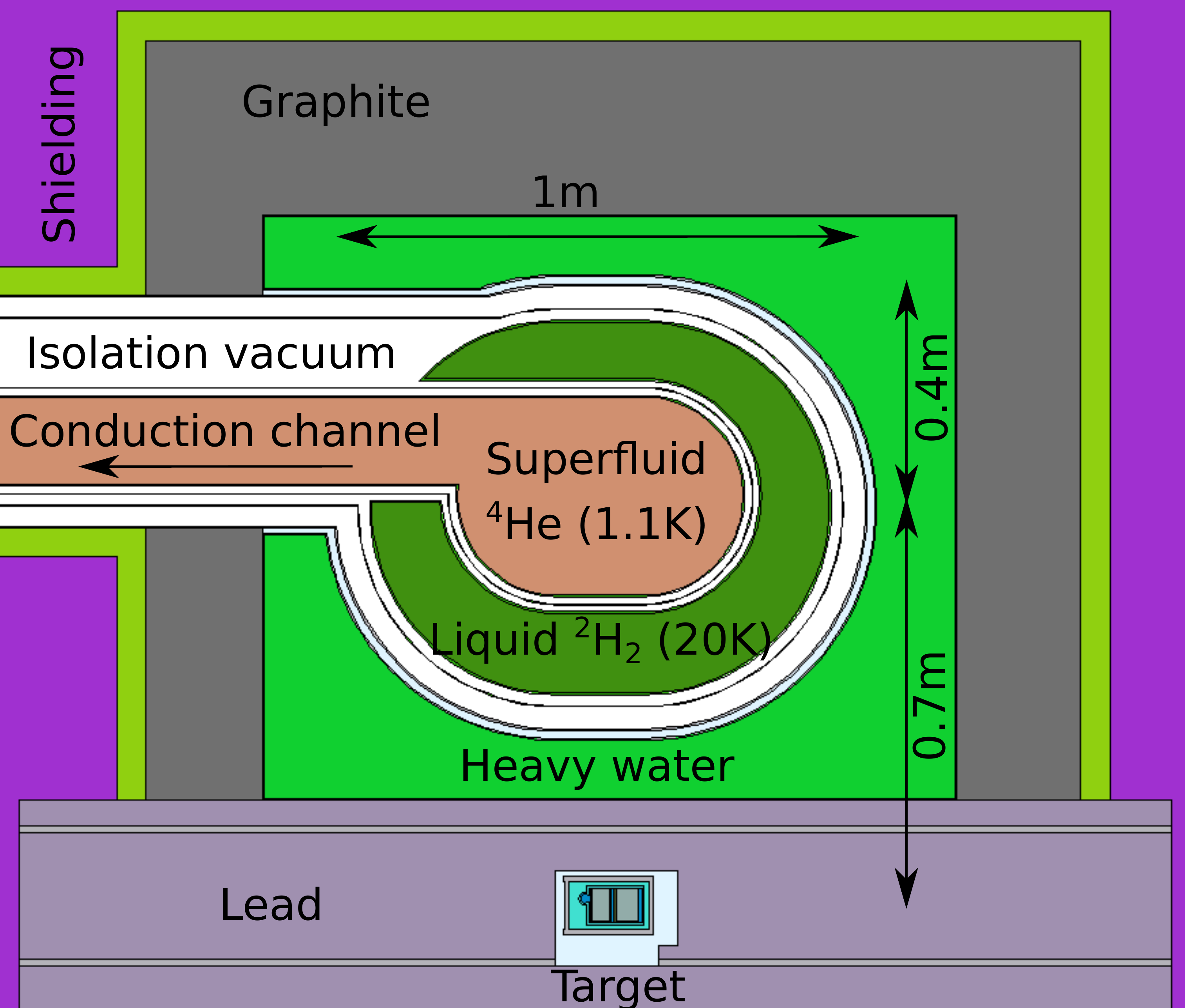}
    \caption{Detailed simulation models of the spallation target (left, top view) and neutron moderators (right, side view). The moderator vessels are made of aluminium 6061.}
    \label{fig:MCNP}
\end{figure}

To estimate UCN production and heat load to the cold moderator and superfluid-helium converter, we performed extensive simulation studies of the spallation target and the neutron moderators with the Monte Carlo software MCNP6.1~\cite{doi:10.13182/NT11-135}. We built a detailed simulation model of the spallation target, taking into account beam windows, target housings, cooling water, gaps in the primary lead moderator, and material compositions determined from assays, see fig.~\ref{fig:MCNP}.


Using a similarly detailed model of the neutron moderators, see fig.~\ref{fig:MCNP}, we performed a multi-parameter optimization, optimizing thicknesses of all moderators layers at once.
The goal of this optimization is to maximize the UCN density delivered to the EDM experiment. A reasonable estimator for the density $\rho$ is the total number of UCN that can be accumulated in the source diluted into the combined volume of source, UCN guides, and experiment $V$:
\begin{equation}
\rho \propto \frac{P \tau}{V}.
\end{equation}
The losses of ultracold neutrons during transport to the experiment should be independent of the source itself and can be neglected for the purpose of this optimization. We performed the optimization for a range of wall lifetimes $\tau_\mathrm{wall}$ between \SIlist{60;100}{\second}, a range of the combined volume of UCN guides and experiment between \SIlist{100;200}{\liter}, and a range of assumptions for the relation between heat load on the converter and its temperature, see section~\ref{sec:heat}.

The optimized result is shown in fig.~\ref{fig:MCNP}. At the full beam power of \SI{20}{\kilo\watt}, the expected UCN-production rate is \SI{2e7}{\per\second}, the expected heat loads on the converter and cold moderator are \SIlist{9.6;38}{\watt}. The converter volume, excluding the UCN guide, contains \SI{34}{\liter} of superfluid helium. North American fire code limits the quantity of liquid deuterium to \SI{150}{\liter}, so the optimizations were performed with a deuterium volume fixed to \SI{125}{\liter}.

With this multi-parameter optimization we were able to do a fair comparison between different cold moderators. Performing the same optimization for solid heavy water and liquid hydrogen showed that liquid deuterium gives a 2.5 times higher UCN density than solid heavy water and a 3 times higher UCN density than liquid hydrogen. Another advantage of liquid deuterium are its well-known neutron-scattering properties already included in MCNP. For solid heavy water there is no specific scattering model available, instead we had to rely on a free-gas model with an effective temperature of \SI{80}{\kelvin}, which is the minimum effective neutron temperature achieved with solid-heavy-water moderators~\cite{doi:10.13182/NSE66-A18558}.

\subsection{Considered improvements}

Using the simulation model, we studied several options to further improve the performance of the source.

The reduced moderation performance of the natural abundance of para-deuterium turned out to reduce performance by at most \SI{5}{\percent}. Hence, we are not including a para-ortho converter in the initial design. The effect of hydrogen contamination in commercially available deuterium also impacts the performance by less than \SI{1}{\percent}.

We considered using a ``neutron filter'' made from cold polycrystalline bismuth between the cold moderator and converter. These simulations were made possible with a neutron-scattering kernel for polycrystalline bismuth provided by Y.-S. Cho~\cite{bismuth_scatteringkernel}. Although the neutron filter increased UCN density by \SI{27}{\percent}, the massive heat load on the cold bismuth and its activation added additional risks and we did not pursue this option further.

Best improvements are achieved by reducing thicknesses of vessels or replacing them with more suitable materials like beryllium. Especially the converter vessel has the highest impact. Replacing it with a vessel made of pure beryllium can increase UCN density in the experiment by \SI{100}{\percent} compared to a vessel made of aluminium 6061. Unfortunately, pure beryllium is prohibitively expensive, but good results can be achieved with beryllium-aluminium alloys like AlBeMet~\cite{AlBeMet} and magnesium-aluminium alloys, which can increase UCN density in the experiment by \SIrange{30}{50}{\percent}.

\section{Cooling superfluid helium to 1\,K}
\label{sec:heat}

\begin{figure}
    \centering
    \includegraphics[width=0.5\textwidth]{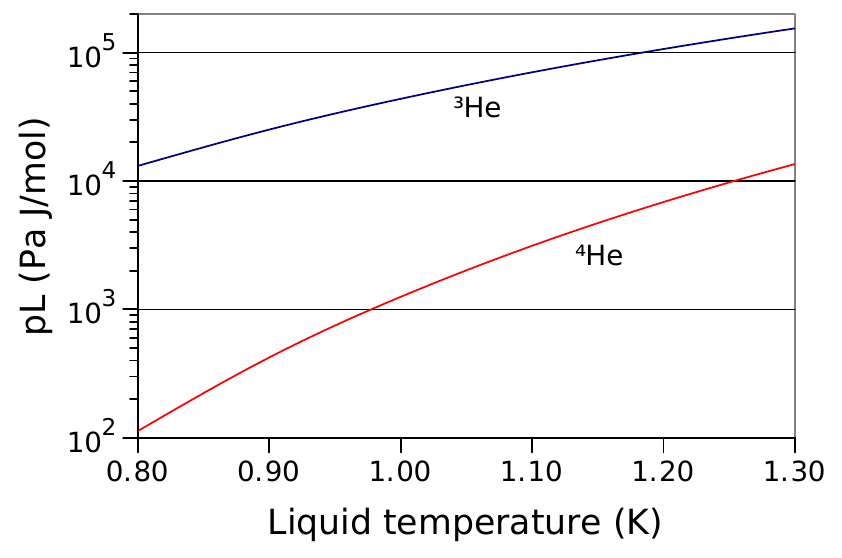}
    \caption{Product of vapor pressure and latent heat of $^3$He and $^4$He \cite{hepak2005}.}
    \label{fig:LxP}
\end{figure}

To cool helium below its boiling point, the vapor pressure above the liquid has to be lowered by pumping. We considered two methods to achieve a helium temperature around \SI{1}{\kelvin}: directly pumping on the superfluid helium itself or pumping on a $^3$He reservoir with a heat exchanger to the superfluid helium.

The heat $Q = \dot{n} L$ removed from a liquid per unit time is given by the molar evaporation rate $\dot{n}$ and the latent heat of vaporization $L$. Assuming the pump pumping on the vapor has a volumetric pumping speed $S$, the evaporation rate is $\dot{n} = p S / (R T_\mathrm{pump})$, where $p$ is the vapor pressure, $R$ the universal gas constant, and $T_\mathrm{pump}$ the vapor temperature at the pump inlet. The cooling power of this pumping system is then
\begin{equation}
Q = \frac{p L S}{R T_\mathrm{pump}}.
\end{equation}

The main advantage of $^3$He is that its product of vapor pressure and latent heat $p L$ in the relevant temperature range is between 10 and 100 times greater than that of $^4$He, see fig. \ref{fig:LxP}. Removing the expected heat load of \SI{10}{\watt} by directly pumping on $^4$He at a temperature of \SI{1.1}{\kelvin} would require a massive pumping speed of \SI{30000}{\cubic\meter\per\hour} or more, depending on the pressure drop in the pumping duct.

\begin{figure}
    \centering
    \includegraphics{f_T_inv.tikz}
    \includegraphics[width=0.4\textwidth]{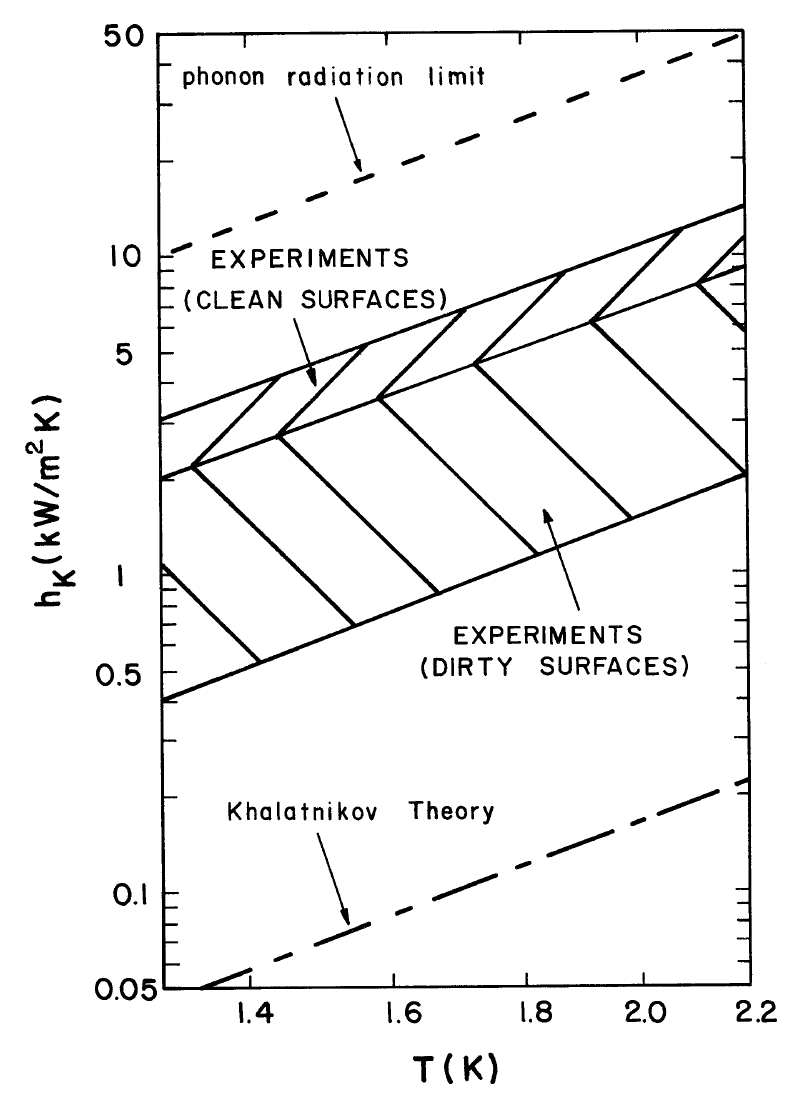}
    \caption{\textit{Left}: Gorter-Mellink heat-conductivity function calculated from different models~\cite{van2012helium,hepak2005,Satoh1988AGMdata}. \textit{Right}: Kapitza conductance over an interface between superfluid helium and copper \cite{van2012helium}.}
    \label{fig:f_t_inv}
\end{figure}

However, $^3$He has an extremely high neutron-absorption cross section and it has to be placed far away from the spallation target. To bridge the distance between the UCN-production volume close to the spallation target and the $^3$He-filled heat exchanger we need a long conduction channel filled with superfluid $^4$He.

The rate of heat transport $Q_\mathrm{GM}$ in a superfluid-helium-filled channel with cross section $A$ and length $L$ from a higher temperature $T_H$ to a lower temperature $T_L$ can be calculated with the Gorter-Mellink model~\cite{GorterMellink1949}:
\begin{equation}
Q_\mathrm{GM} = \left( \frac{A^3}{L} \int_{T_L}^{T_H} f(T)^{-1} dT \right)^{1/3}.
\end{equation}
The heat-conductivity function $f(T)^{-1}$ has been experimentally determined with high accuracy at temperatures above \SI{1.4}{\kelvin}. At lower temperatures, however, different models deviate from each other, see fig.~\ref{fig:f_t_inv}.


The interfaces between superfluid helium, heat exchanger, and $^3$He introduce additional resistance to heat transport. The heat transport from a liquid with temperature $T_l$ to a solid with temperature $T_s$ is $Q_K = h_K (T_s - T_l)$. The so-called Kapitza conductance $h_K$ is typically estimated with the Khalatnikov model~\cite{van2012helium} multiplied with an empirical scaling factor $k_G$:
\begin{equation}
h_K = k_G \cdot \SI{20}{\watt\per\square\meter\per\kelvin\tothe{4}} \cdot T_l^3.
\end{equation}
The Kapitza conductance between superfluid helium and copper has been well studied. It depends largely on surface quality and the scaling factor $k_G$ varies between 10 for dirty surfaces and 80 for clean surfaces, see fig.~\ref{fig:f_t_inv}. Taking into account the density and sound velocity of $^3$He, we can estimate the Kapitza conductance between copper and $^3$He from the Khalatnikov model, giving $h_{K (^4\mathrm{He})}/h_{K (^3\mathrm{He})} = 1.2 \sim 2.6$. At KEK, we are currently performing measurements of heat conduction in superfluid helium and Kapitza conductance to validate these calculations. First results are encouraging and point to a $k_G$ of around \num{40}.

To be able to fit sufficient radiation shielding between the target and the $^3$He cryostat, the conduction channel needs to be about \SI{2}{\meter} long. Gorter-Mellink conductance strongly depends on the cross section of the channel, so we chose a large diameter of \SI{15}{\centi\meter}. The conduction channel also acts as UCN guide (see section \ref{sec:transport}), i.e. the surface of the heat exchanger in contact with the superfluid has to have a UCN-compatible, smooth surface. This smooth surfaces cannot be made arbitrarily large, since it would also increase the required amount of expensive $^3$He. Hence, the Kapitza conductance of the superfluid-copper interface incurs the largest temperature gradient.

Assuming we can cool the $^3$He reservoir to \SI{0.8}{\kelvin}, removing the expected heat load of \SI{10}{\watt} with a pumping speed on the order of \SI{10000}{\cubic\meter\per\hour}, we can calculate the temperature gradients caused by Kapitza and Gorter-Mellink conductance. With a $k_G$ of 40, the expected temperature of the UCN-production volume is between \SIlist{1.08;1.14}{\kelvin}. Kapitza conductance accounts for two thirds of the temperature difference between $^3$He and UCN-production volume, Gorter-Mellink conductance for one third.

When taking into account all these effects, the temperatures reached with a direct-pumping solution and a $^3$He fridge are comparable. The cost for the additional pumping speed required for a direct-pumping solution are comparable to the cost of $^3$He and additional isotopically purified $^4$He required for a $^3$He fridge. Ultimately, we chose a $^3$He fridge since it can reach lower temperatures at smaller heat loads, while the rapidly dropping $^4$He vapor pressure limits the direct-pumping solution to temperatures above \SI{1.1}{\kelvin}. Additionally, the direct-pumping performance is critically dependent on the difficult-to-estimate pressure drop in the pumping duct; the higher $^3$He vapor pressure allows more flexibility in the duct design.

\section{Transport of ultracold neutrons}
\label{sec:transport}

\begin{figure}
    \centering
    \includegraphics[width=\textwidth]{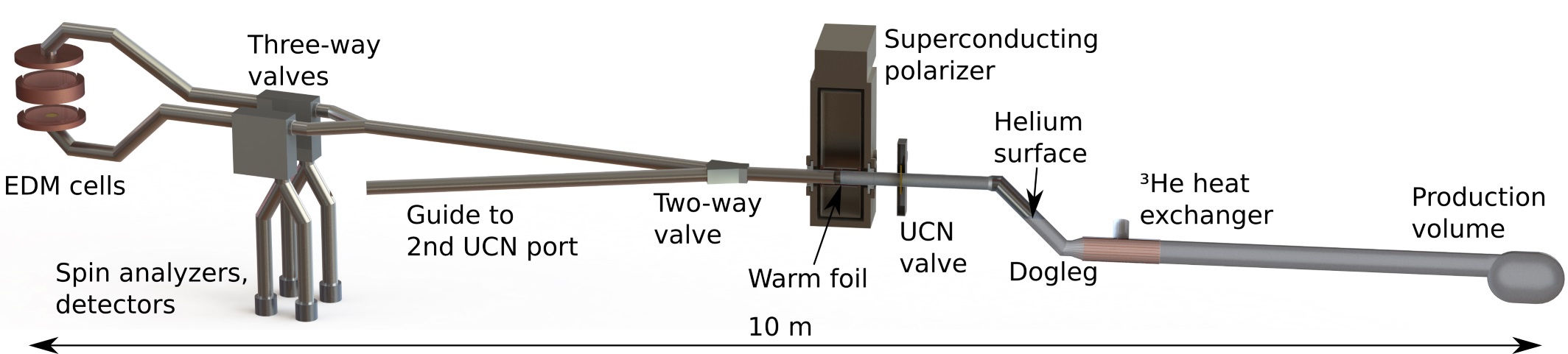}
    \caption{UCN-guide geometry and UCN-handling components for the next-generation UCN source and EDM experiment. Neutron moderators, radiation shielding, $^3$He cryostat, and magnetic shielding for the EDM experiment are not shown.}
    \label{fig:UCNtransport}
\end{figure}

The ultracold neutrons produced in the source have to be transported from the production volume buried in radiation shielding to the EDM experiment or a planned second UCN port. About \SI{10}{\meter} of UCN guides are required, see fig. \ref{fig:UCNtransport}. Their transport efficiency is one of the most critical parameters determining the UCN density that can be delivered to the experiment. To optimize the guide geometry, we performed extensive analytical and simulation studies.

The diameter of the superfluid-helium-filled UCN guide needs to be large, since it doubles as a heat-conduction channel (see section \ref{sec:heat}). Although large diameters reduce UCN-transport losses, the increased volume dilutes the UCN density in the EDM experiment. Studies with varying diameters, taking into account the simultaneously varying temperature gradients, show that a diameter of \SIrange{15}{18}{\centi\meter} is optimal. Even larger diameters are favorable for the heat exchanger. However, due to the costs of larger amounts of $^3$He required for a larger heat exchanger we chose diameters for conduction channel and heat exchanger of \SI{15}{\centi\meter}. To reduce heat load, the UCN production volume and conduction channel have aluminium-6061 walls. Since the Fermi potential of aluminium is too small to confine useful numbers of ultracold neutrons, it will be plated with nickel-phosphorus, allowing the source to store ultracold neutrons with energies up to \SI{213}{\nano\electronvolt}.

We decided to gravitationally confine the superfluid helium by adding a vertical dogleg to the UCN guide downstream of the heat exchanger, see fig. \ref{fig:UCNtransport}. This design avoids a cold window separating the superfluid helium from the guide vacuum. A warm window is still required to avoid contamination of the source, but there is no risk of freezing residual gas on the cold window, reducing UCN transmission. Increasing the height of the dogleg will soften the UCN spectrum, allowing longer storage times in the experiment. However, a higher dogleg will lower the high-energy cutoff of the spectrum. Our studies currently favor a height between \SIlist{30;70}{\centi\meter} for the EDM experiment.

Above the dogleg, the UCN guide transitions from cryogenic to room temperature and ends in a UCN valve followed by a warm  foil separating the helium vapor from the UCN-guide vacuum. The valve will allow different operation modes, e.g. a batch mode accumulating ultracold neutrons in the source and then opening the valve, or a steady-state mode continuously producing ultracold neutrons with the valve open. The foil will most likely be an \SI{0.1}{\milli\meter}-thick aluminium foil. To avoid losing all ultracold neutrons with energies below the Fermi potential of aluminium, we will place the superconducting polarizer magnet required for the EDM experiment around the foil, see fig. \ref{fig:UCNtransport}. We are considering other foil materials with lower Fermi potentials, like titanium or polyethylene, which would give us more flexibility where to place the polarizer.

To avoid diluting the UCN density, the UCN guides downstream of the heat exchanger will have a smaller diameter than the conduction channel. A narrower dogleg additionally reduces heat load due to thermal radiation, conduction, convection, and superfluid film flow in the transition region. An existing superconducting polarizer magnet limits the guide diameter to \SI{8.5}{\centi\meter}. However, our studies seem to favor a slightly larger diameter of \SI{10}{\centi\meter}, potentially requiring a new polarizer. Downstream of the polarizer, the guides have to be coated with materials preserving the UCN polarization, e.g. nickel-molybdenum or nickel-phosphorus. A two-way valve will provide ultracold neutrons to either the EDM experiment or the second experimental port.

The final leg of the UCN guides penetrates through the radiation shielding and splits to supply ultracold neutrons to both storage cells of the EDM experiment, see fig. \ref{fig:UCNtransport}. To estimate how long the EDM experiment has to operate to achieve the aspired statistical sensitivity, we performed studies taking into account all the aforementioned effects: temperature-dependent up-scattering in superfluid helium and helium vapor, losses on reflection of UCNs from walls, shifting of the spectrum in the dogleg, losses in the valves, and losses in the foil and polarizer. Similar simulation studies for our current prototype source showed excellent agreement with the experimental data~\cite{Ahmed:2018fvb}. For the EDM experiment, we also take into account polarization losses during transport and during storage in the experiment, spectrum-dependent storage times in the experiment, and losses during UCN detection.

The most important parameter is the transport efficiency of the UCN guides. We typically assume a diffuse-reflection probability of \SI{3}{\percent}, giving a transmission efficiency for straight guides of \SI{90}{\percent\per\meter}. With these assumptions we expect a polarized-UCN density in the EDM cells of \SI{350}{\per\cubic\centi\meter}. Taking into account typical time periods during which the EDM experiment cannot operate---e.g. during working hours with increased environmental noise, while magnetic shields are de-gaussed, while the electric field is inverted, etc.---we expect to reach a sensitivity of \SI{1e-27}{\elementarycharge\centi\meter} within 400 beam days.

\section{Conclusions}

We worked out a conceptual design for a next-generation source for ultracold neutrons at a new, dedicated neutron spallation target at TRIUMF. It will be based on a superfluid-helium converter, cooled to \SI{1.1}{\kelvin} by a $^3$He fridge, and a liquid-deuterium cold-neutron moderator. Detailed design of neutron moderators, the liquid-deuterium system, $^3$He cryostat, and radiation shielding is currently ongoing and supported by extensive studies to optimize it for a neutron-EDM experiment. The source is scheduled for installation and commissioning in 2021.

For the new EDM experiment, we are currently testing first components, including high-voltage electrodes with co-magnetometer gases, magnetometers, and magnetic-field coils. The recently installed prototype UCN source helps us to characterize UCN guides, storage cells, valves, polarizers, and detectors.


\bibliographystyle{unsrt}
\bibliography{bib}

\begin{thebibliography}{10}

\bibitem{0038-5670-34-5-A08}
Andrei~D Sakharov.
\newblock {Violation of CP invariance, C asymmetry, and baryon asymmetry of the
  universe}.
\newblock {\em Soviet Physics Uspekhi}, 34(5):392, 1991.

\bibitem{PhysRevLett.13.138}
J.~H. Christenson, J.~W. Cronin, V.~L. Fitch, and R.~Turlay.
\newblock {Evidence for the $2\ensuremath{\pi}$ Decay of the $K_{2}^{0}$
  Meson}.
\newblock {\em Phys. Rev. Lett.}, 13:138--140, Jul 1964.

\bibitem{PhysRevLett.87.091801}
B.~Aubert et~al.
\newblock {Observation of $\mathit{CP}$ Violation in the ${\mathit{B}}^{0}$
  Meson System}.
\newblock {\em Phys. Rev. Lett.}, 87:091801, Aug 2001.

\bibitem{PhysRevD.95.096014}
Francesco Capozzi, Eleonora Di~Valentino, Eligio Lisi, Antonio Marrone,
  Alessandro Melchiorri, and Antonio Palazzo.
\newblock Global constraints on absolute neutrino masses and their ordering.
\newblock {\em Phys. Rev. D}, 95:096014, May 2017.

\bibitem{POSPELOV2005119}
Maxim Pospelov and Adam Ritz.
\newblock Electric dipole moments as probes of new physics.
\newblock {\em Annals of Physics}, 318(1):119 -- 169, 2005.
\newblock Special Issue.

\bibitem{PhysRev.78.695}
Norman~F. Ramsey.
\newblock A molecular beam resonance method with separated oscillating fields.
\newblock {\em Phys. Rev.}, 78:695--699, Jun 1950.

\bibitem{PhysRevD.92.092003}
J.~M. Pendlebury et~al.
\newblock Revised experimental upper limit on the electric dipole moment of the
  neutron.
\newblock {\em Phys. Rev. D}, 92:092003, Nov 2015.

\bibitem{PhysRevC.95.045503}
G.~Bison et~al.
\newblock Comparison of ultracold neutron sources for fundamental physics
  measurements.
\newblock {\em Phys. Rev. C}, 95:045503, Apr 2017.

\bibitem{GOLUB1975133}
R.~Golub and J.M. Pendlebury.
\newblock Super-thermal sources of ultra-cold neutrons.
\newblock {\em Physics Letters A}, 53(2):133 -- 135, 1975.

\bibitem{PhysRevC.97.012501}
T.~M. Ito et~al.
\newblock {Performance of the upgraded ultracold neutron source at Los Alamos
  National Laboratory and its implication for a possible neutron electric
  dipole moment experiment}.
\newblock {\em Phys. Rev. C}, 97:012501, Jan 2018.

\bibitem{LAUSS201498}
Bernhard Lauss.
\newblock {Ultracold Neutron Production at the Second Spallation Target of the
  Paul Scherrer Institute}.
\newblock {\em Physics Procedia}, 51:98 -- 101, 2014.
\newblock ESS Science Symposium on Neutron Particle Physics at Long Pulse
  Spallation Sources, NPPatLPS 2013.

\bibitem{Kahlenberg2017}
J.~Kahlenberg, D.~Ries, K.~U. Ross, C.~Siemensen, M.~Beck, C.~Geppert, W.~Heil,
  N.~Hild, J.~Karch, S.~Karpuk, F.~Kories, M.~Kretschmer, B.~Lauss, T.~Reich,
  Y.~Sobolev, and N.~Trautmann.
\newblock {Upgrade of the ultracold neutron source at the pulsed reactor TRIGA
  Mainz}.
\newblock {\em The European Physical Journal A}, 53(11):226, Nov 2017.

\bibitem{PhysRevC.90.015501}
F.~M. Piegsa, M.~Fertl, S.~N. Ivanov, M.~Kreuz, K.~K.~H. Leung,
  P.~Schmidt-Wellenburg, T.~Soldner, and O.~Zimmer.
\newblock {New source for ultracold neutrons at the Institut Laue-Langevin}.
\newblock {\em Phys. Rev. C}, 90:015501, Jul 2014.

\bibitem{PhysRevLett.108.134801}
Yasuhiro Masuda, Kichiji Hatanaka, Sun-Chan Jeong, Shinsuke Kawasaki, Ryohei
  Matsumiya, Kensaku Matsuta, Mototsugu Mihara, and Yutaka Watanabe.
\newblock {Spallation Ultracold Neutron Source of Superfluid Helium below 1 K}.
\newblock {\em Phys. Rev. Lett.}, 108:134801, Mar 2012.

\bibitem{Ahmed:2018fvb}
S.~Ahmed et~al.
\newblock {First ultracold neutrons produced at TRIUMF}.
\newblock {\em Submitted to Phys. Rev. C}, 2018.

\bibitem{Yu1986}
Z-Ch. Yu, S.~S. Malik, and R.~Golub.
\newblock A thin film source of ultra-cold neutrons.
\newblock {\em Zeitschrift f{\"u}r Physik B Condensed Matter}, 62(2):137--142,
  Jun 1986.

\bibitem{GOLUB1979387}
R.~Golub.
\newblock {On the storage of neutrons in superfluid 4He}.
\newblock {\em Physics Letters A}, 72(4):387 -- 390, 1979.

\bibitem{PhysRevC.92.024004}
P.~Schmidt-Wellenburg, J.~Bossy, E.~Farhi, M.~Fertl, K.~K.~H. Leung, A.~Rahli,
  T.~Soldner, and O.~Zimmer.
\newblock Experimental study of ultracold neutron production in pressurized
  superfluid helium.
\newblock {\em Phys. Rev. C}, 92:024004, Aug 2015.

\bibitem{KOROBKINA2002462}
E.~Korobkina, R.~Golub, B.W. Wehring, and A.R. Young.
\newblock {Production of UCN by downscattering in superfluid He4}.
\newblock {\em Physics Letters A}, 301(5):462 -- 469, 2002.

\bibitem{PhysRevC.93.025501}
K.~K.~H. Leung, S.~Ivanov, F.~M. Piegsa, M.~Simson, and O.~Zimmer.
\newblock {Ultracold-neutron production and up-scattering in superfluid helium
  between 1.1 K and 2.4 K}.
\newblock {\em Phys. Rev. C}, 93:025501, Feb 2016.

\bibitem{PDG2018}
M.~Tanabashi et~al.
\newblock {Review of Particle Physics}.
\newblock {\em Phys. Rev. D}, 98:030001, 2018.

\bibitem{BL1U}
S.~Ahmed et~al.
\newblock {A new beamline for fundamental neutron physics at TRIUMF}.
\newblock {\em To be published in Nucl. Instrum. Methods Phys. Res. A}, 2018.

\bibitem{doi:10.13182/NT11-135}
T.~Goorley et~al.
\newblock {Initial MCNP6 Release Overview}.
\newblock {\em Nuclear Technology}, 180(3):298--315, 2012.

\bibitem{doi:10.13182/NSE66-A18558}
J.~J. Rush, D.~W. Connor, and R.~S. Carter.
\newblock {Study of D2O Ice as a Cold-Neutron Source}.
\newblock {\em Nuclear Science and Engineering}, 25(4):383--389, 1966.

\bibitem{bismuth_scatteringkernel}
Young-Sik Cho and Jonghwa Chang.
\newblock The calculation of neutron scattering cross sections for silicon and
  bismuth crystal at thermal energies.
\newblock {\em Journal of Nuclear Science and Technology}, 39(sup2):176--179,
  2002.

\bibitem{AlBeMet}
Materion Corporation.
\newblock {\em AlBeMet Technical Fact Sheet}.

\bibitem{hepak2005}
Cryodata.Inc.
\newblock {\em {HEPAK} User's Guide}.
\newblock Horizon Technologies, 2005.

\bibitem{van2012helium}
Steven~W Van~Sciver.
\newblock {\em Helium cryogenics}.
\newblock Springer Science \& Business Media, 2012.

\bibitem{Satoh1988AGMdata}
Mineo Okuyama, Toshimi Satoh, and Takeo Satoh.
\newblock {Adiabatic flow of He II: Motion of normal fluid component and
  vortices}.
\newblock {\em Physica B: Condensed Matter}, 154(1):116 -- 124, 1988.

\bibitem{GorterMellink1949}
C.~J. Gorter and J.~H. Mellink.
\newblock {On the irreversible processes in liquid helium II}.
\newblock {\em Physica}, 15(3):285 -- 304, 1949.

\end{thebibliography}

\end{document}